\documentclass[
    ,final            
  ]
  {aipproc}

\layoutstyle{6x9}

\begin{document}

\title{Observations of X-ray Flares from Gamma Ray Bursts}


\author{A.~D.~Falcone}{
  address={Pennsylvania State University, 525 Davey Lab, University Park, PA 16802}
}
\author{D.~Morris}{
  address={Pennsylvania State University, 525 Davey Lab, University Park, PA 16802}
}
\author{J.~Racusin}{
  address={Pennsylvania State University, 525 Davey Lab, University Park, PA 16802}
}
\author{G.~Chincarini}{
  address={INAF -- Osservatorio Astronomico di Brera, Merate, Italy},
  altaddress={Universit\`a degli studi di Milano-Bicocca, Dipartimento di Fisica, Milano, Italy}
}
\author{A.~Moretti}{
  address={INAF -- Osservatorio Astronomico di Brera, Merate, Italy}
}
\author{P.~Romano}{
  address={INAF -- Osservatorio Astronomico di Brera, Merate, Italy}
}
\author{D.~N.~Burrows}{
  address={Pennsylvania State University, 525 Davey Lab, University Park, PA 16802}
}
\author{C.~Pagani}{
  address={Pennsylvania State University, 525 Davey Lab, University Park, PA 16802}
}
\author{M.~Stroh}{
  address={Pennsylvania State University, 525 Davey Lab, University Park, PA 16802}
}
\author{D.~Grupe}{
  address={Pennsylvania State University, 525 Davey Lab, University Park, PA 16802}
}
\author{S.~Campana}{
  address={INAF -- Osservatorio Astronomico di Brera, Merate, Italy}
}
\author{S.~Covino}{
  address={INAF -- Osservatorio Astronomico di Brera, Merate, Italy}
}
\author{G.~Tagliaferri}{
  address={INAF -- Osservatorio Astronomico di Brera, Merate, Italy}
}
\author{N.~Gehrels}{
  address={NASA/Goddard Space Flight Center, Greenbelt, MD 20771}
}

\begin{abstract}
{\it{Swift}}-XRT observations of the X-ray emission from gamma ray bursts (GRBs) and during the GRB afterglow have led to many new results during the past two years. One of these exciting results is that $\sim$$1/3-1/2$ of GRBs contain detectable X-ray flares. The mean fluence of the X-ray flares is $\sim$10$\times$ less than that of the initial prompt emission, but in some cases the flare is as energetic as the prompt emission itself. The flares display fast rises and decays, and they sometimes occur at very late times relative to the prompt emission (sometimes as late as 10$^5$ s after T$_0$) with very high peak fluxes relative to the underlying afterglow decay that has clearly begun prior to some flares. The temporal and spectral properties of the flares are found to favor models in which flares arise due to the same GRB internal engine processes that spawned the prompt GRB emission. Therefore, both long and short GRB internal engine models must be capable of producing high fluences in the X-ray band at very late times.
 
\end{abstract}

\maketitle


\section{Introduction}
Since its launch on 2004 November 20, {\it{Swift}} \citep{geh04} has provided detailed measurements of numerous gamma ray bursts (GRBs) and their afterglows with unprecedented reaction times. By detecting burst afterglows promptly, and with high sensitivity, the properties of the early afterglow and extended prompt emission can be studied in detail for the first time. This also facilitates studies of the transition between the prompt emission and the afterglow. The rapid response of the pointed X-ray Telescope (XRT) instrument \cite{bur05a} on {\it{Swift}} has led to the discovery that large X-ray flares are common in GRBs and occur at times well after the initial prompt emission. 

While there are still many unknown factors related to the mechanisms that produce GRB emission, the most commonly accepted model is that of a relativistically expanding fireball with associated internal and external shocks \citep{mes97}. In this model, internal shocks produce the prompt GRB emission. Observationally, this emission typically has a timescale of $\sim30$ s for long bursts and $\sim$0.3 s for short bursts \citep{mee96}. The expanding fireball then shocks the ambient material to produce a broadband afterglow that decays quickly (typically as ${\propto}t^{-\alpha}$). When the Doppler boosting angle of this decelerating fireball exceeds the opening angle of the jet into which it is expanding, then a steepening of the light curve (jet break) is also predicted \citep{rho99}. For a description of the theoretical models of GRB emission and associated observational properties, see \cite{mes02, zha04, piran05, van00, der99}.

With the advent of recent {\it{Swift}}-XRT observations of many large flares at various times after the burst, it is clear that a new constraint on GRB models is available. Two pre-{\it{Swift}} observations of relatively small flux increases \cite{pir05} did not provide a complete picture of the X-ray flaring activity, and where not initially interpreted as flares; instead they were interpreted as afterglow onsets. Recent observations by XRT indicate that flares are common, that they can have a fluence comparable to the initial prompt emission, and that they have various timescales, spectra, and relative flux increase factors \cite{bur05b, fal06, romano06, bur07}. By studying the properties of these flares, we may elucidate the nature of both the X-ray flares and the GRB internal engine.

These GRB flare studies can take two forms; individual GRBs with flares can be studied in detail or a large sample of flares can be studied to look at global properties of flares. The former approach has been taken by many authors \cite{fal06,romano06,bur05b,mor06,goa07,pag06,cus07,zha06}. The latter method of looking at large samples is used in four recent papers \cite{fal07, chi07, koc07, but07}. 

\section{Selection of Remarkable Flaring GRBs}
\begin{figure}
\centering
\includegraphics[width=0.4\linewidth]{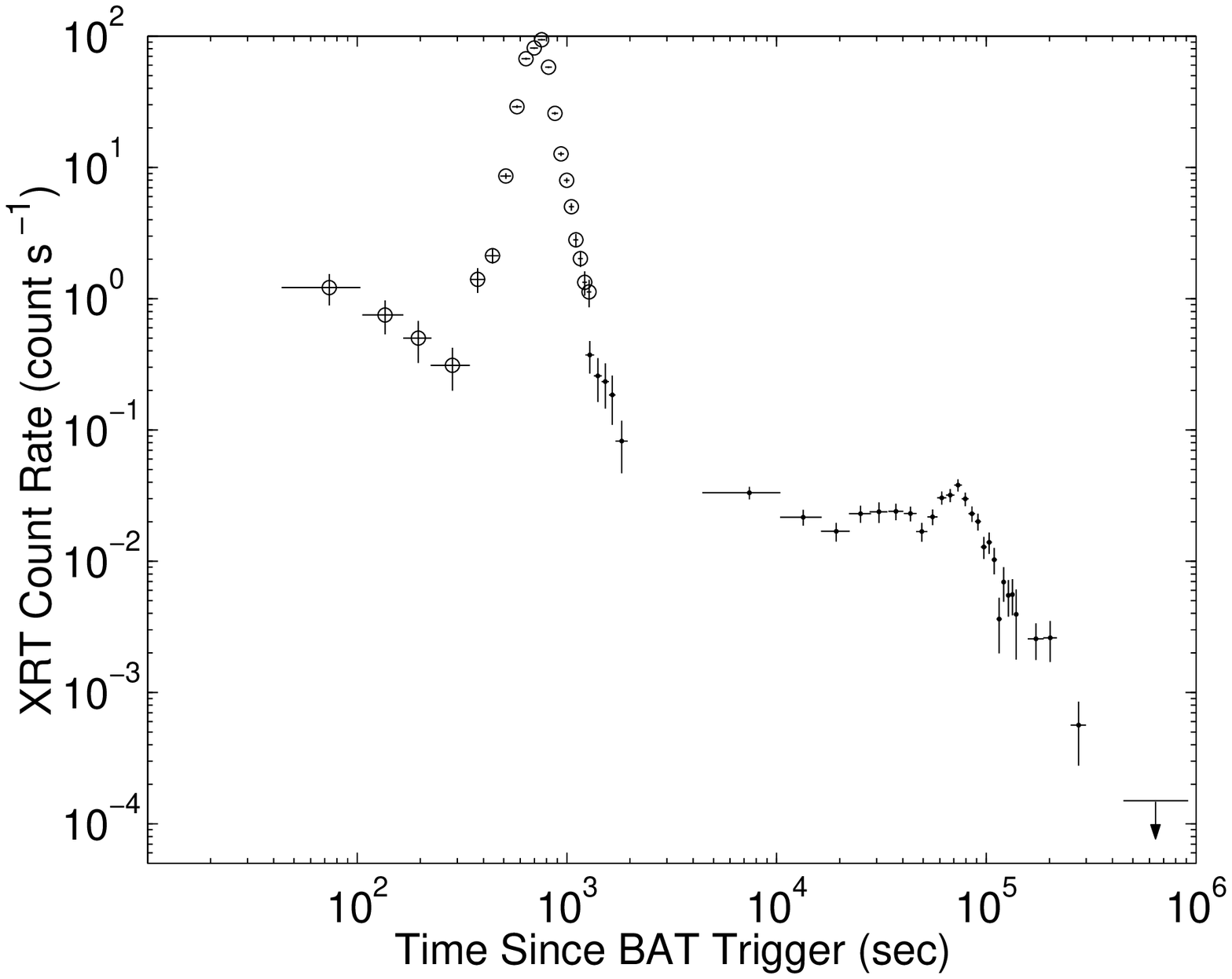}\quad
\includegraphics[width=0.45\linewidth,clip,bb=95 389 457 630]{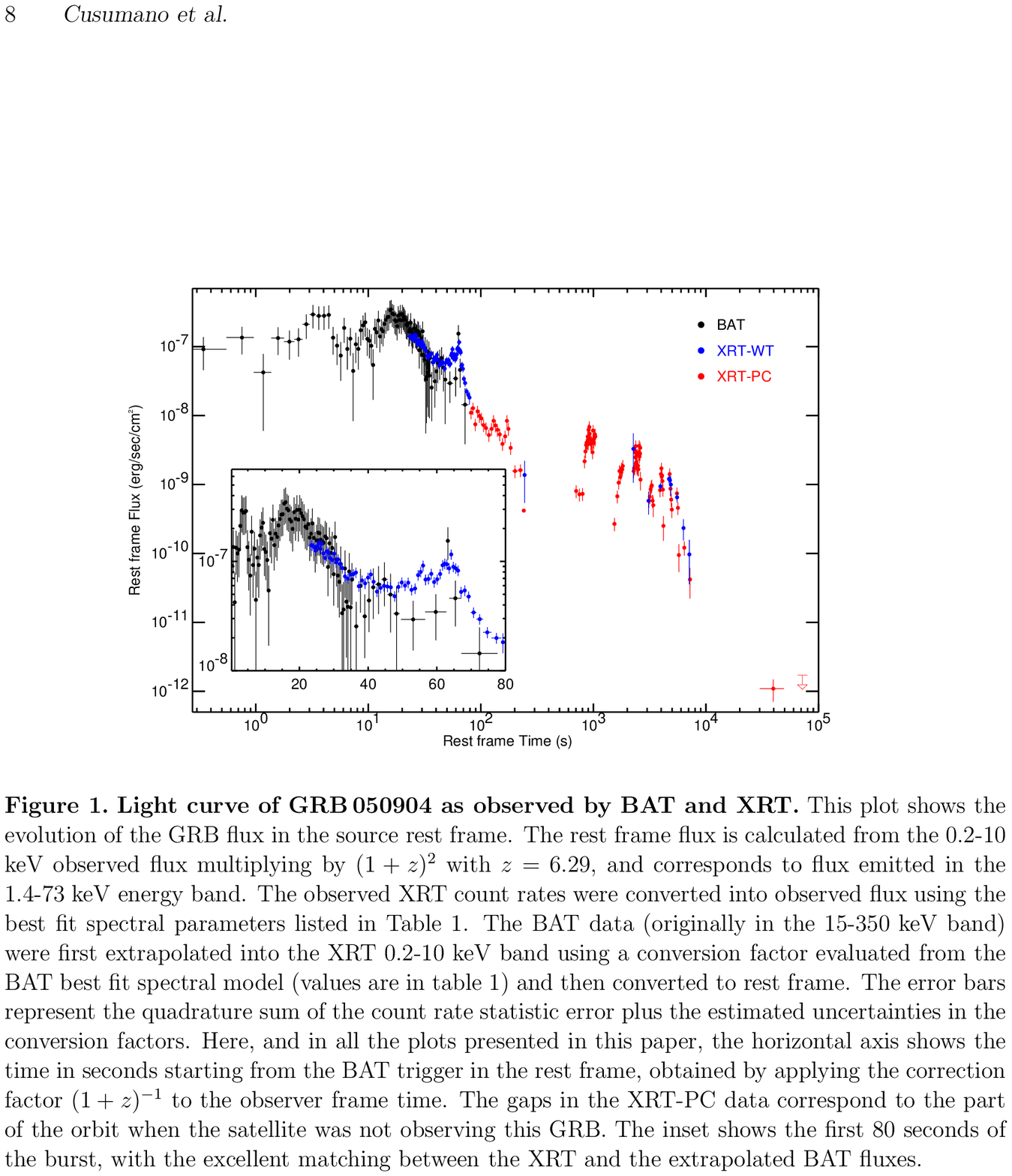}
\caption{(LEFT) X-ray light curve of GRB 050502B. For details, see \citet{fal06}.  (RIGHT) X-ray light curve of GRB~050904, transformed into the rest frame for z=6.29. The BAT data points are from an extrapolation into the XRT energy range (black points, from 0-75~s) superimposed on the XRT light curve (20-10,000~s). For details, see \citet{cus07}.}
\label{fig:050502B_050904_lc}
\end{figure}

GRB 050502B is a prime example of a GRB with at least one large flare at late times after the cessation of the initial prompt emission detected by BAT \cite{fal06, bur05b}. The X-ray light curve is shown in Figure~\ref{fig:050502B_050904_lc}. A giant flare, with a flux increase by a factor of $\sim$500, was observed. The fluence during the giant flare, (1.2 $\pm$ 0.05) $\times 10^{-6}$ erg cm$^{-2}$ in the 0.2-10 keV band, was greater than or equal to the prompt fluence. The flare rises to a sharp peak at 743 $\pm$ 10 s. In the hard band (1-10 keV), there is significant short time structure within the peak of the giant flare itself. The spectrum can be fit best by an absorbed cutoff power law (or Band function) \cite{ban93}, rather than a simple absorbed power law,  which fits the underlying afterglow nicely. For details, see \citet{fal06}. The spectral index hardens significantly during the flare (with a cutoff energy of $\sim$2.5 keV in the XRT band) before returning back to a softer and more typical afterglow spectrum following the flare. Before and after the flare, the temporal decay of the underlying afterglow can be fit well with a single power law $\sim t^{-0.8 \pm 0.2}$. At much later times, between (1.9 $\pm$ 0.3) $\times10^{4}$ s and (1.1 $\pm$ 0.1) $\times10^{5}$ s, there are two broad bumps (or possibly one broad bump with some structure). GRB 060526 has a flare that is remarkably similar to the giant flare of GRB 050502b, with a flux increase of about $100\times$ relative to the underlying afterglow. In this case, the giant flare clearly consists of two separate, overlapping flares, peaking at about 220 s and 300 s. These giant flares dominate the light curve and resemble the prompt GRB, although at lower peak energies and usually longer time-scales.

From studies of a few individual GRBs with several strong flares, temporal evolution of spectral properties has been explored. Spectral evolution in individual bright flares has been seen in GRB 050406 \cite{romano06}, GRB 050502B \cite{fal06}, GRB 050607 \cite{pag06}, GRB 060714 \cite{mor06}, GRB 050822 \cite{god07}, as well as several others. In all cases, spectral hardening was observed at the onset of the flare, followed by spectral softening as the flare peaked and decayed. Temporal evolution of spectra from flare to flare has also been seen, as in GRB 060714 which shows a decrease in $E_{peak}$ as a function of flare onset time \cite{kri07}. \citet{but07} have observed the same trend in their study of several bright X-ray flares.

The exceptional short bursts, GRB 050724 and GRB 070724, exhibited significant flaring detected by XRT \cite{bar05a, cam06, gru06, zia07}. It is also possible that GRB 051227, which has a significant X-ray flare peaking at $\sim110$ s, is a short burst \cite{bar05b}. However, there is some ambiguity in its characterization as short or long. It is certainly clear that the short burst and the long burst mechanisms must both be capable of producing X-ray flares.

GRB 050904, at a redshift of 6.29, is the most distant GRB detected to date. This burst has a very interesting X-ray light curve (see Figure~\ref{fig:050502B_050904_lc}) with many flares superimposed on top of the underlying temporal decay, and on top of one another \cite{cus07}. Even after a transformation of the light curve into the rest frame of the GRB, there is significant flaring at times as late as $\sim$5000 s.

\section{A Large Sample of Flares}
While the detailed study of individual flares is important, it is equally important to look at the properties of the flares in a more general sense to look for trends and overall mean properties of the flares and to compare these properties to those of the typical prompt and afterglow emission. One such flare sample was chosen by looking at all Swift light curves, between launch and 2006 January 24, and eliminating the ones that did not show any hint of deviation from a power law decay with typical breaks. Only the flares with S/N$>3$ were retained in the sample. This analysis results in 77 flares in 33 GRBs. Although some GRBs had $>$7 flares, the average was 2 flares per GRB. The temporal and spectral properties of this flare sample have been studied by \citet{fal07} and \citet{chi07}, and some results are briefly described below.

\subsection{Spectral Properties and Fluence of Sample}
The spectral parameters of the flares were fit in a way that accounted for the contribution from the underlying light curve photons, which can have a significant impact. This is, of course, made difficult by the fact that the canonical light curve has a variety of breaks \citep{nou06} and may be contaminated by additional flares. In general, simple absorbed power law spectra provide a reasonable fit to the flare data from the sample, but more complex models such as Band functions typically provide a better fit along with more self consistent parameters (particularly N$_H$, as discussed in \citep{fal06}).
\begin{figure}
\centering
\includegraphics[width=0.4\linewidth,clip,bb=290 185 60 390,angle=180]{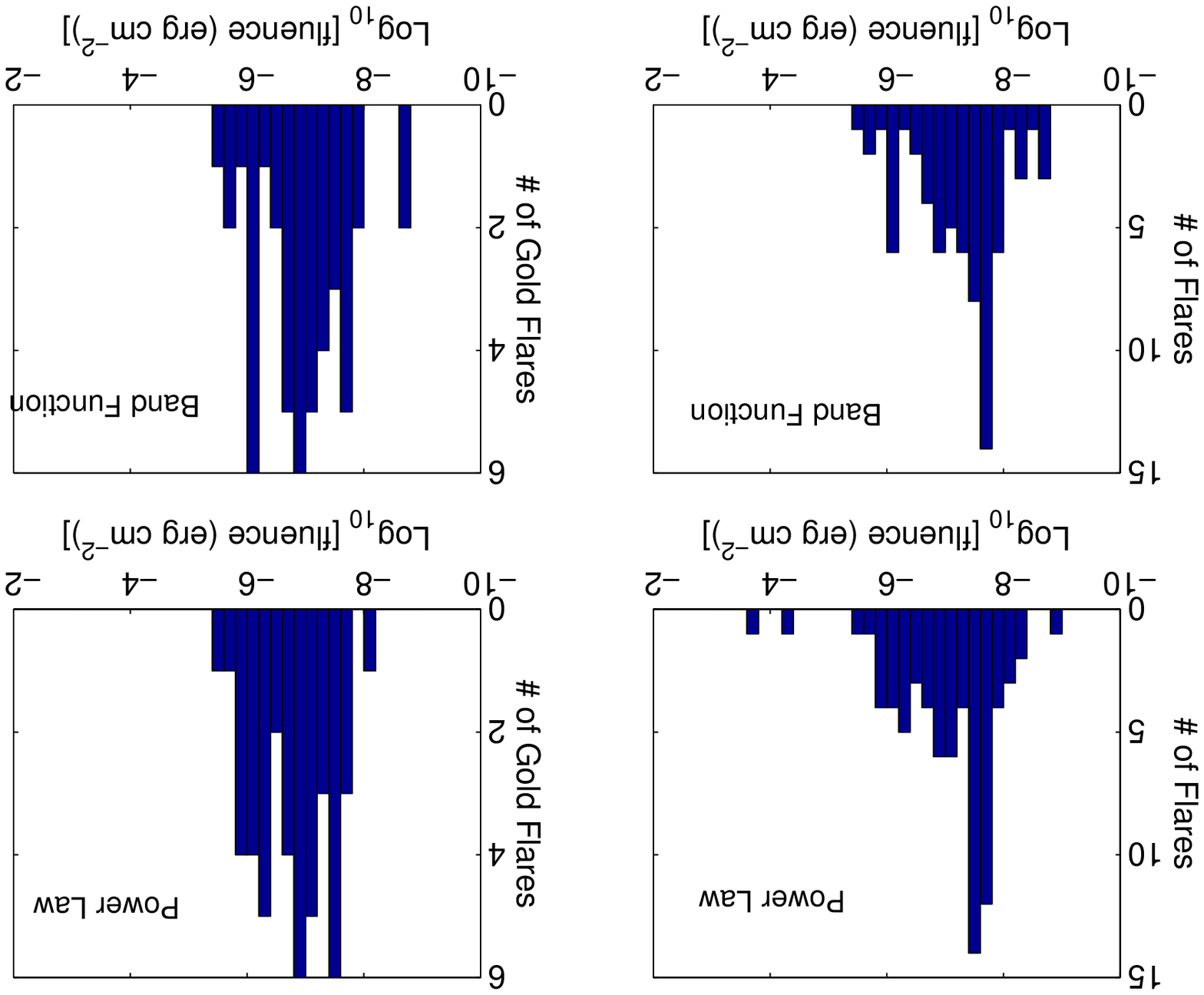}\quad
\includegraphics[width=0.45\linewidth]{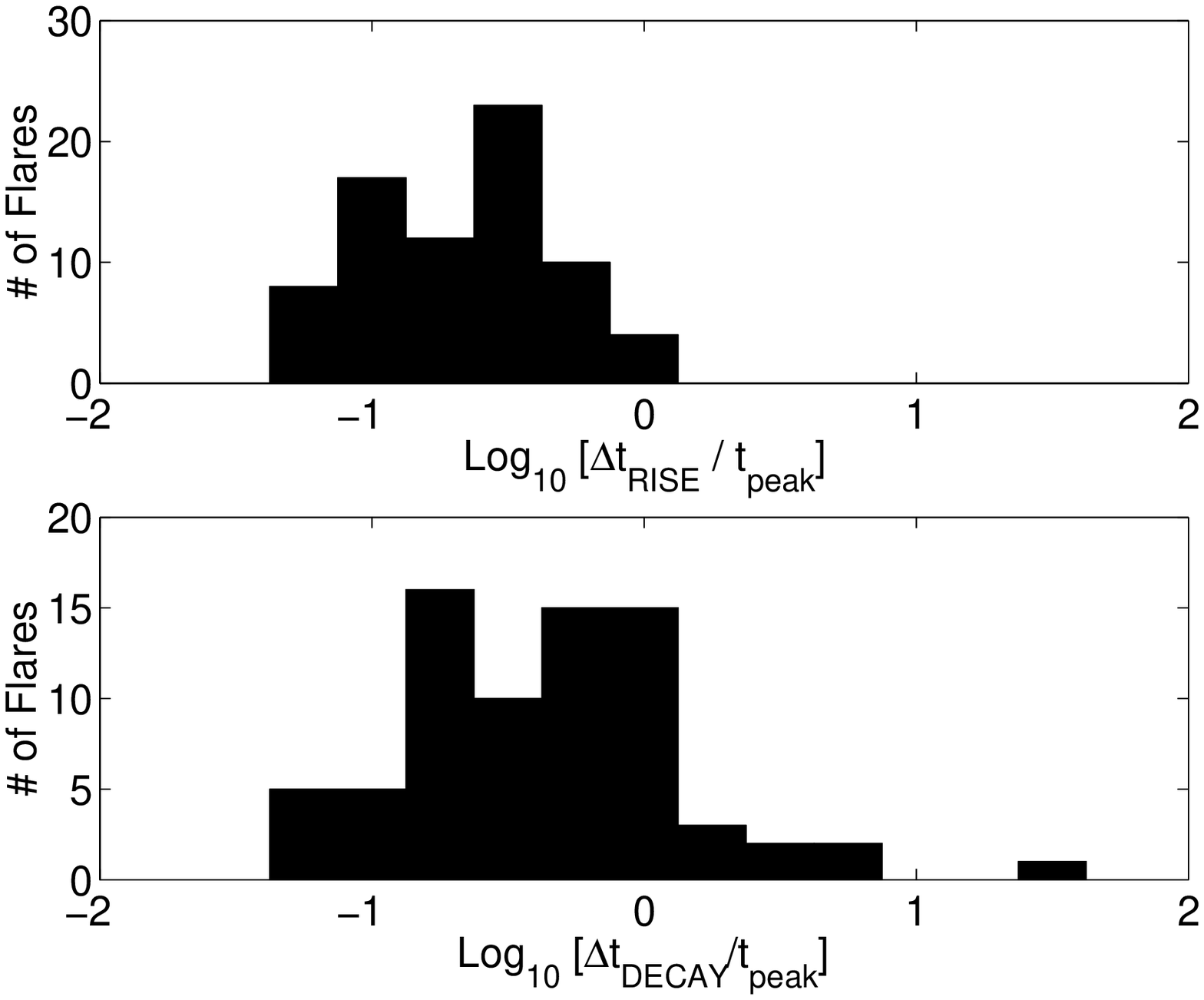}
\caption{(LEFT) Unabsorbed 0.2--10.0 keV fluence distribution of 77 flares derived from Band function fits. (RIGHT) Distributions of rise and decay times for flares in our sample.}
\label{fig:fluence_temporal}
\end{figure}
The overall distribution of flare fluences are shown in Figure \ref{fig:fluence_temporal}. This fluence value does not include the contribution from the power law component of the spectral model that was used to approximate the underlying afterglow contribution to the flare spectrum. The frequently used practice of quoting the total fluence, without subtracting the underlying afterglow, is misleading since the underlying afterglow light curve sometimes contributes a large fraction of the total fluence. The mean 0.2-10.0 keV fluence (unabsorbed) of our sample of flares, derived using Band function fits, is $2.4\times10^{-7}$ erg cm$^{-2}$, with a standard deviation of $5.3\times10^{-7}$ erg cm$^{-2}$.

\subsection{Temporal properties of Sample}
Each flare light curve in the sample of 77 flares was fit with a model consisting of an underlying power law decay with a flare superimposed on it. The rising portion of the flare and its decay were fit with power laws. We then define $\Delta t_{rise}$ as
$t_p - t_{start}$, where $t_p$ is the peak time (the time when the
rising power law intersects the decaying power law), and $t_{start}$ is the
time when the rising power law intersects the underlying afterglow's
power law decay. Similarly, the decay time is defined as $\Delta t_{decay} = t_{stop}-t_p$.
Figure~\ref{fig:fluence_temporal} shows the distributions of
rise and decay times for our sample. The distributions peak near $\Delta t/t \sim 0.3$, and there are several flares with $\Delta t/t < 0.1$. Since external shock processes are typically much slower than this \citep{iok05,nak06}, a faster process such as internal shocks is a more likely explanation for the flares (see \cite{der08} for alternative view). The temporal properties of the sample is described in more detail by \citet{chi07}.

\section{Conclusions}
Based on a sample drawn from the first 110 {\it{Swift}} GRBs, it is clear that significant X-ray flares are produced frequently and at late times. For some GRBs, there have been more than 7 significant flares measured. For some flares, the fluence in the 0.2-10.0 keV X-ray band is in excess of the prompt emission fluence in the 15-150 keV band. The rise and decay times of flares are generally very fast relative to the peak times of the flares. Frequently, it is clear that a nominal afterglow light curve decay has begun prior to a flare and that it resumes following the flare. The spectra during flares typically shows an increase in the hardness ratio that gradually declines back to the pre-flare value as the flare flux decays. The spectra of flares are frequently fit better with Band functions (normally used to describe prompt emission from GRBs) than with simple power laws (normally used for afterglows). Based on this list of properties, it is likely that the origin of flares is more akin to the origin of prompt GRB emission than it is to afterglow emission. Within the context of the standard model, this would require late time internal shocks.

The initial flare sample contained 14 GRBs with a measured redshift, and the average redshift did not differ from the average redshift for all {\it{Swift}} GRBs, including those without flares. This implies that late flares can not be merely explained as redshifted normal prompt emission. This result is also supported by individual burst analysis that are corrected for redshift, such as that of \citet{cus07} in which a high redshift burst is corrected by (1+z) and still has very late flares. All of this implies that the large fluence values for flares reported in this paper (sometimes comparable to the nominal prompt emission) must be produced at very late times with peak energies in the X-ray band. GRB progenitor models must be capable of producing this late emission within the same energy budget that was previously applied to only the nominal prompt emission. The fallback of material onto the central black hole after a stellar collapse could last for long time periods \cite{woo93, mac01} and lead to late internal engine activity, but the reduced luminosity expected from this model at late times eliminates it as an explanation for all flares. Several models for continued activity of the central engine have been proposed (e.g. \cite{per06,fan06,dai06,kin05,pro05,kat97}). These models, and others, must be evaluated within the context of the energy budget and the measured spectral parameters.

During X-ray flares, the excess flux of X-ray photons at late times may provide a previously unknown seed population of photons that can be inverse Compton scattered. If these are observed at high energies, the Lorentz factor could be constrained by comparing the X-ray synchrotron peak and the Inverse Compton peak. Therefore, X-ray flares may provide a new mechanism by which GLAST and very high energy gamma-ray instruments can study GRBs.

\end{document}